\documentclass[fleqn,10pt]{wlscirep}

\title{Reduction of power grid fluctuations by communication between smart devices}

\author[1,a]{Eder Batista Tchawou Tchuisseu}
\author[1,b]{Dami\`a Gomila}
\author[1,c]{Pere Colet}
\affil[1]{Instituto de F\'{\i}sica Interdisciplinar y Sistemas Complejos, IFISC (CSIC-UIB), Campus Universitat Illes Balears, 07122 
Palma de Mallorca, Spain}

\affil[a]{ederbatista@ifisc.uib-csic.es}
\affil[b]{damia@ifisc.uib-csic.es}
\affil[c]{pere@ifisc.uib-csic.es}

\begin{abstract}

The increase of electric demand and the progressive integration of renewable sources threatens the stability of the power grid. To solve this issue, several methods have been proposed to control the demand side instead of increasing the spinning reserve in the supply side. Here we focus on dynamic demand control (DDC), a method in which appliances can delay its scheduled operation if the electric frequency is outside a suitable range. We have recently shown that DDC effectively reduces small and medium size frequency fluctuations but, due to the need of recovering pending tasks, the probability of large demand peaks, and hence large frequency fluctuations, may actually increase. Although these events are very rare they can potentially trigger a failure of the system and therefore strategies to avoid them have to be addressed. In this work we introduce a new method including communication among DDC devices belonging to a given group, such that they can coordinate opposite actions to keep the group demand more stable. We show that for this method the amount of pending tasks decreases by a factor 10 while large frequency fluctuations are significantly reduced or even completely avoided.
\end{abstract}
\begin{document}

\flushbottom
\maketitle
%
%


\section*{Introduction}

The stable operation of the electric power grid is based on several control methods and suitable management that includes demand forecasting, technical maintenance, generation and transmission planning, etc. The control of  the grid frequency is one of the most important tasks of the system operator to ensure global stability, reliability and efficiency of the grid operation. 
Indeed, the frequency control is strongly related to balance supply (power generation) and demand (load/power consumption). If at any moment supply surpasses the load, then the frequency rises, while if load exceeds supply, the frequency falls. This balance is not easy to achieve
because of the random character of (part of) the load and the increasing supply fluctuations due to the integration of renewable energy sources. Possible power line outages or power plant failures can introduce additional isolated large fluctuations. Thus, in face of a power unbalance, traditional operation modifies the supply side to follow the load (demand). This requires power plants to have extra generation capacity (spinning reserve) allowing to adjust the output power according to frequency variations \cite{Rebours2007,Short07,Dehghanpour}. Besides, typical grids include also a non-spinning or supplemental reserve integrated by fast-response generation units that can be brought online in a few minutes.

The intensive integration of renewable energy sources (solar photovoltaic capacity alone is expected to increase by 43\% by 2022 according to the US Energy Information Agency) and the constant increase in electricity demand from residential buildings, which accounts for about 30-40 \% of the total energy use all over the world \cite{haider2016review,alberini2011,esther2016},
have increased the difficulty to manage the frequency of the system with the traditional philosophy. This has motivated electric company utilities to adopt a new philosophy which tries to force the demand to follow the supply. This approach, known as demand side management (DSM) aims to distribute the control of the power system among the users \cite{esther2016}. DSM includes a number of methods to control the energy consumption at the customer side of the meter \cite{Mohsenian}. This way of controlling the grid has the potential to significantly reduce the expensive spinning and supplemental reserve needs leading to beneficial economical effects to both utility companies and customers \cite{Takagi,Moghadam,Saadat,Shi}.

Different DSM methods have been proposed \cite{Lu2006,Albadi2008,US_DEPARTMENT,Albadi2007}, most of them designed for commercial, industrial and big customers such as Universities \cite{Alvarez,Escriva}. For instance, G. Escriv\'a-Escriv\'a \textit{et al.} applied a demand-response program to about $1500$ Heater, Ventilators and Air-Conditioners (HVAC) spread around the campus of the Universidad Polit\'ecnica de Valencia \cite{Escriva}. They compared different control strategies by proposing different schedules for a common HVAC, and found a good strategy optimizing energy and economic savings. Fewer demand-response programs have been implemented for residential customers. Some propose using technologies such as smart appliances and smart meters combined to a bidirectional communication between a Load Serving Entity (LSE) and the smart appliances, or among a home energy management system and both a LSE and smart appliances \cite{Shi,Dehghanpour,Shu,Ruiz}. Indeed in \cite{Shi} Shi \textit{et al.} formulated a residential demand-response program as an optimal power flow problem, and proposed a distributed scheme where the LSE and the households interactively communicate to compute an optimal demand schedule. In \cite{Mohsenian}, Amir-Hamed \textit{et al.} presented an autonomous and distributed DSM system based on game-theoretic energy consumption scheduling where users interacting through a communication system are players and their strategies are implemented through the daily schedules of their households. It was shown that the global optimal scenario in terms of minimizing the energy costs is obtained at the Nash equilibrium of the formulated game.

In general DSM needs smarts meters for communication, such that the LSE can send a real time price signal which is essential to implement such programs \cite{Albadi2008,Shi,Qian,Joe}. If the frequency decreases the price rises, inducing customers to switch off appliances. Conversely, if the frequency rises the price decreases encouraging customers to switch on appliances.
Finally, instead of using price as input in DSM one can just use the frequency, which is universal and readily available at any point in the grid. This is the idea behind Dynamic Demand Control (DDC) which uses algorithms integrated into an appliances such as HVAC or refrigerators to regulate its energy consumption according to the instantaneous frequency. One of the advantages of this control is the direct connection with the grid. In \cite{Short07}, Short \textit{et al.} applied DDC to a population of $1000$ refrigerators by modifying the temperature thresholds of the thermostat control system of the refrigerator. The thermostat was tweaked such that the two temperature limits are modified by an amount proportional to the real time frequency deviation from the nominal value. This technology can be incorporated in different electrical appliances to manage peak loads and to balance second-to-second supply and demand   \cite{Schweppe,Lasseter1,Lasseter2,Lemoine,Doherty,Nehrir,Brooks,Shively,Takagi,Molderink,Kosek,Parliament,Nick,National, Hirst,Trevor,Hild,Moghadam,Luo}. 

In \cite{Tchawou1} we introduced a dynamical model, including power generation and fluctuating demand, to study the general effects of DDC on the grid frequency. We find that while DDC is effective in reducing small and medium size fluctuations, the need to recover pending tasks changes the probability distributions of the fluctuations introducing large tails so that the probability of rare events in the form of large frequency fluctuations becomes non negligible. This collateral effect of DDC poses a risk to the system and it can potentially trigger a failure of the power grid. Strategies to avoid such undesired effects have then to be addressed. Here we show that introducing communication between smart devices is a convenient strategy to significantly reduce or even suppress large frequency fluctuations. The communication among devices belonging to given group or cluster allows them to coordinate switchings such that the power consumption of the cluster is kept as constant as possible. As a consequence, the number of accumulated pending tasks as well as the overall frequency fluctuations are reduced improving the performance of DDC. 

\section*{Model}

We consider the simple model introduced in \cite{Tchawou1} consisting of a set of differential equations describing the dynamics of a power plant, a stochastic demand model and a generic DDC protocol. The dynamics of the power plant is described by the well-known swing equation for the frequency of the system $\omega$ together with primary and secondary regulation \cite{Saadat,Tchawou1}:
\begin{eqnarray}
\frac{d\omega}{dt} &=& \frac{\omega}{2HP_{\rm G}}(P_{\rm m}-P_{\rm e}) \label{eq1} \\
\frac{dP_{\rm m}}{dt} &=& \frac{1}{\tau_g}[P_{s}-P_{\rm m}-\frac{P_{\rm G}}{R\omega_{\rm R}}(\omega-\omega_{\rm R})] \label{eq2} 
\\
\frac{dP_{s}}{dt} &=& -\frac{K}{\omega_{\rm R}}(\omega-\omega_{\rm R})
\label{eq3} 
\end{eqnarray} 
where  $\omega_{\rm R}$ is the reference rotation frequency, 
$P_{\rm m}$ is the generated mechanical power, and $P_{\rm e}$ is the total electric power. $P_{\rm G}$ and $H$ are the nominal capacity and the inertia constant of the generator.
Under a supply-load unbalance, the governor, described by Eqs.~(\ref{eq2},\ref{eq3}), is responsible to restore the frequency to its reference value, and it does so in two steps. First primary control, Eq.~(\ref{eq2}), increases (decreases) $P_{\rm m}$ to halt a frequency decline (rise). Then, at a slower time scale, secondary regulation, Eq.~(\ref{eq3}), restores the frequency to its reference value. Here $P_{\rm s}$ is the spinning reserve power, $K$ is the gain of the secondary controller, and $\tau_g$ is the characteristic time of the power plant. 

The total electric load $P_{\rm e}$ can be divided in two parts: a non frequency-sensitive load and a frequency-sensitive load (as for example that associated to electrical motors) so that 
such that
\begin{equation}  
P_{\rm e}(\omega,t)= \left(1+ D \frac{\omega-\omega_{\rm R}}{\omega_{\rm R}}\right) P(t),
\label{Pe} 
\end{equation}
where $D$ is proportional to the fraction of the load which is frequency sensitive and $P(t)$ is the load at $\omega=\omega_{\rm R}$. Here we take $D=0.026$.

The time dependent load $P(t)$ is modeled\cite{Tchawou1} by considering a population of $N$ appliances (or bunch of devices that operate together). For simplicity we assume appliances can only be in two states: off, in which there is no consumption and on, in which the appliance have the rated power $P_0$ when operating at $\omega=\omega_{\rm R}$. We also assume that at an appliance in the off state can be powered on with probability rate $p$, such that $pdt$ is the probability that the off-state appliance is turned on in the time interval $dt$. Similarly we assume that an appliance in the on state can be switched off with probability rate $q$. For $p=q$ in the stationary regime typically half of the devices are on and half off. Thus, the time dependent load of appliance $j$, is given by
\begin{equation}  
 P_j(t) = \left \{ \begin{array}{ll} P_{0}, & {\rm on~state~at~time~} t \\ 0, & {\rm off~state~at~time~} t \end{array} \right .
 \label{Pj} 
 \end{equation}
The total load of the system is given by the addition of the load of the $N$ appliances: 
\begin{equation}  
P(t)=\sum\limits_{\textit{j=1}}^\textit{N} P_{j}(t),
\label{Ptotal} 
\end{equation} 
 
Throughout this paper we take the load parameters $N=1000$ and $p=q=6.55 \times 10^{-4}$ and the power plant parameters $\omega_{\rm R}=50$Hz, $H=2.26$ s, $\tau_g=0.78$, $R=0.07$, $P_{\rm G}/P_0=N/2$ and $K/P_0=50$ s$^{-1}$. Results are independent of $P_0$.

DDC is implemented on top of the time dependent load. The smart appliances following this protocol make their switchings conditional on the instantaneous value of the frequency. This process somehow provides an effective energy storage by accumulating pending tasks, namely by delaying the switchings of the devices. Indeed, DDC is conceived such that a smart appliance that has to switch on, it does so only if the frequency is above a defined threshold $\omega > \omega_R-\epsilon$. Otherwise it remains off saving energy. This saved energy is accounted as pending tasks ($Q$). Similarly a smart appliance that has to switch off, does so only if $\omega>\omega_R+\epsilon$. Otherwise it remains on consuming extra energy, which is also accounted as pending tasks.   
Smart devices with pending tasks monitor the frequency continuously. Energy-consuming pending tasks can only be recovered when the frequency is above a second threshold,  $\omega > \omega_R+\epsilon_1$, while energy-saving pending tasks can only be recovered when $\omega<\omega_R-\epsilon_1$. Smart devices with pending tasks monitor the frequency continuously and once the frequency is suitable, actual recovery of a task takes place randomly with probability $\gamma$ to avoid simultaneous switching-on of several devices.  

In \cite{Tchawou1} we determined proper values of the parameters $\epsilon_1$ and $\gamma$ to achieve efficient DDC operation. In particular, $\epsilon$=$0.05$, $\epsilon_1=0.06$ and $\gamma=1,2 \times 10^{-3}$ yielded a good reduction of frequency fluctuations without accumulating too many pending tasks. These values are to be used here.

\section*{Results}

To analyze the performance of the control we consider the absolute value of the frequency fluctuations $\Delta \omega (t)= |\omega (t) - \omega_{\rm R}|$ and determine the complementary cumulative distribution $R(\Delta \omega)$ which corresponds to the probability of having a frequency fluctuation with absolute value larger than $\Delta \omega$.
From numerical simulations we obtain a set of values $\{\Delta \omega_k \}$ at discrete times $t_k$, $k=1,...,M$ which is then reordered from the smallest to the largest value $\{\Delta \omega_i \}$, so that $i$ is the rank of $\Delta \omega_i$, $i=1,...,M$. Then the complementary cumulative distribution is estimated as
\begin{equation} 
R(\Delta \omega_i) =1-\frac{i-1}{M-1}. 
\end{equation}

\begin{figure}
\centering
\includegraphics[width=0.6\linewidth]{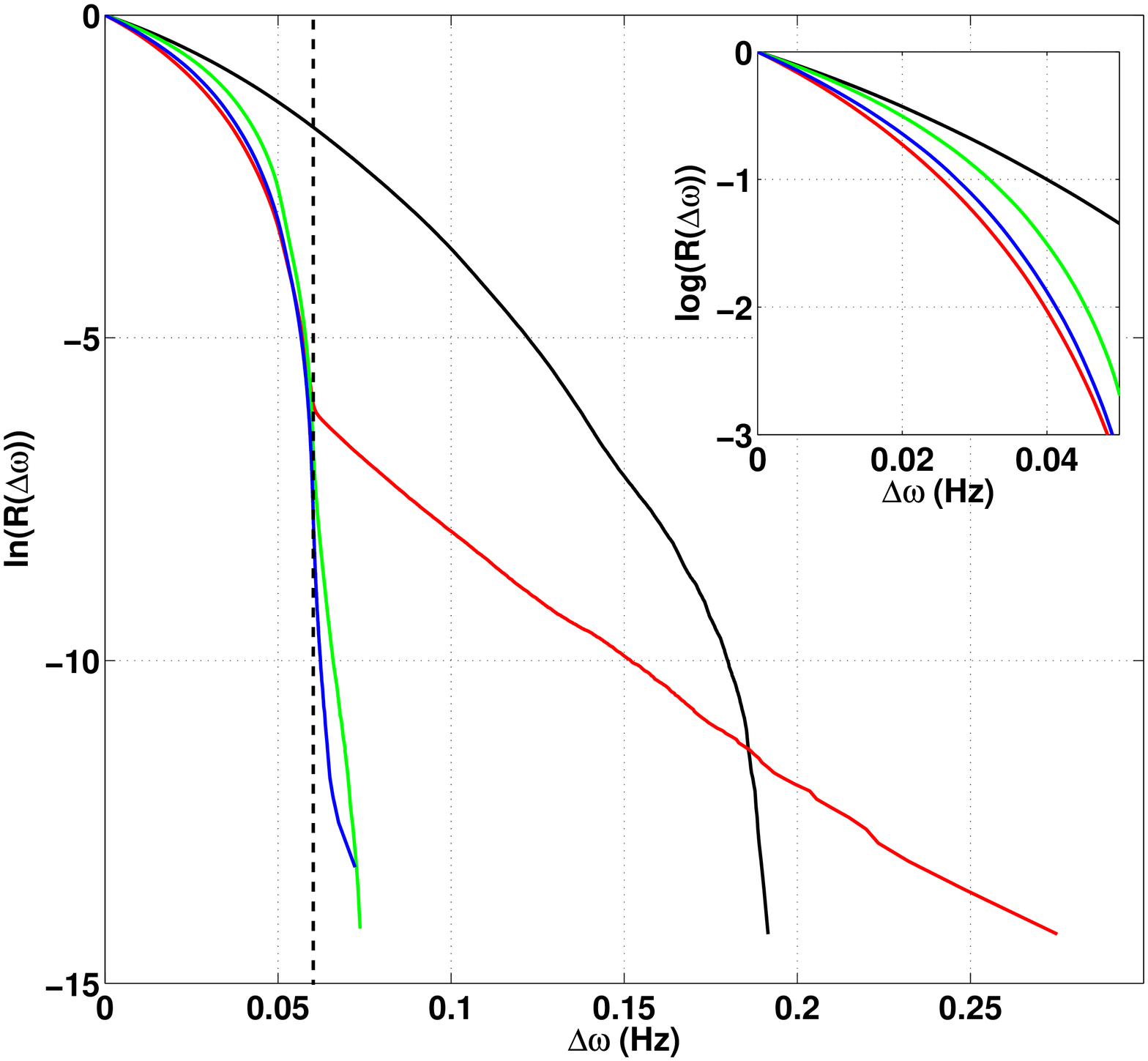}
\caption{Estimated complementary cumulative distribution $R(\Delta \omega)$ which indicates the probability of having a frequency fluctuation whose absolute value is larger than $\Delta\omega$=$|\omega-\omega_{\rm R}|$. The black line corresponds to the case in which no DDC is used as obtained from a numerical simulation with $N=1000$ devices. The red line shows $R(\Delta \omega)$ when applying the basic DDC protocol to the same system. Green and blue lines show the results when DDC devices communicate among them (CeDDC). The green line corresponds to the case of all-to-all communication while the blue line corresponds to the case where communication takes place within clusters of $250$ devices. The inset shows in detail the initial part of the curves.} 
\label{cumulative_distribution} 
\end{figure}

Figure \ref{cumulative_distribution} shows the estimated values for $R(\Delta \omega)$ obtained from a numerical simulation with $1000$ devices. Black symbols correspond to the case in which no DDC is applied while red symbols show the results for the same system when all devices apply the basic DDC protocol introduced in \cite{Tchawou1}. Small and medium size fluctuations are significantly reduced and the variance of the fluctuations decreases. However the shape of the distribution changes, it now there is a crossover at $\Delta \omega \approx \epsilon_1$ beyond which the complementary cumulative distribution decays exponentially (linearly in the logarithmic scale of Fig.~ \ref{cumulative_distribution}). As a consequence the probability of large-scale fluctuations is, in fact, larger than without DDC. The change in the shape of the complementary cumulative distribution comes from the constant necessity of recovering pending tasks in the DDC basic protocol, which triggers large demand peaks that translate in large frequency fluctuations, potentially harmful for real power grids. This effect can not be really removed tweaking parameters. For this reason, in this work we propose introducing direct interactions among appliances as an strategy to recover pending tasks without triggering large frequency fluctuations, enhancing the efficiency of DDC.

We consider that smart devices belonging to a given group or cluster can communicate among them. The group could be, for instance, composed by the smart devices belonging to the same household. The interaction protocol proposed is as follows: when a device has to switch on or off, either due to a user action or to recover pending tasks, but the frequency conditions do not allow it, the device will check if another device in the same group has performed the opposite action during a past time window of duration $T$. If so, the device is allowed to switch on or off independently of the value of the frequency. As a consequence, this protocol allows more switchings of individual devices than without interaction, reducing the number of pending tasks, while the power consumption by each group has smaller variations. We note that this protocol does not intend to bring the frequency back to the reference value as soon as possible, but rather aims at reducing fluctuations at the group scale, giving time to the primary and secondary regulation to operate. We refer to this protocol as communication-enhanced DDC (CeDDC).

Smart devices can communicate though a hub such as a household energy management in which case the hub has to keep an accounting of the devices that have switched on or off in the past time $T$. It is also possible to implement this protocol with direct communication among devices as follows: Each smart device keeps a {\em power released} register $\mathcal{P}$ which can take three values: $+1$, $0$ and $-1$. At switching on (off) the smart device sets $\mathcal{P} =-1$ ($\mathcal{P} =+1$). After a time window $T$ the device resets $\mathcal{P}=0$. The idea is that $\mathcal{P}=+1$ ($\mathcal{P}-1$) potentially allows another device to be powered on (off) without changing the overall consumption of the cluster.
Let $j$ be a smart device that has to turn on (resulting of a user action or to recover pending tasks) but the frequency is out of range to do so. Device $j$ searches for other devices in his group (in a given order order or randomly) until it finds one, say device $k$ such that $\mathcal{P}_k=+1$. Then $j$ stops the search, powers on and sets $\mathcal{P}_j=-1$ while device $k$ resets $\mathcal{P}_k=0$ since it is no longer capable to provide a slot for other devices to be powered on. 

Although in here we only consider the case where all devices have the same rated power $P_0$, this protocol can be generalized to devices with different rated power or with a non-constant consumption by using a register $\mathcal{P}$ which instead of storing $+1$ or $-1$ at switching, it stores the actual value of the power released at the switching time. This value is positive if the device switches off and negative if it switches on. This is the amount of energy made available to other devices that intend to perform the opposite operation during a time slot $T$. A device $j$ which has to switch but the frequency is out of range would search for other devices in his cluster until finding one, $k$, that it covers its needs and  
$\mathcal{P}_k$ is adjusted to account for the power used by device $j$. It is also possible to consider more sophisticated procedures in which device $j$ combines the power made available by several devices, although in this case implementation is easier using a centralized hub.

Figure \ref{cumulative_distribution} displays the results for the complementary cumulative distribution $R(\Delta \omega)$ when communication among smart devices is included (CeDDC), for two cases, when all devices communicate with each other (green symbols) and when grouped in $4$ clusters of size $250$ (blue symbols). It can be clearly seen  that, in addition to a large reduction of the frequency fluctuations variance, the long tails of the probability distribution are also largely suppressed.

\begin{figure}
\centering
   \includegraphics[width=0.6\linewidth] {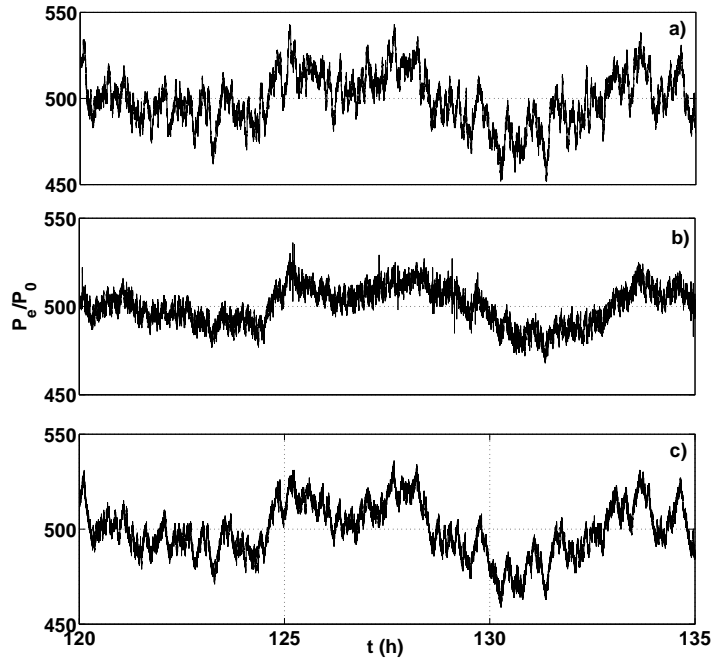}
 \caption{Time evolution of the scaled total electric load $P_{\rm e}/P_0$:  without DDC (a), with the basic DDC protocol (b) and with CeDDC when all devices communicate with each other (c).}
\label{load_trajectory}
\end{figure}

\begin{figure}
\centering
   \includegraphics[width=0.6\linewidth] {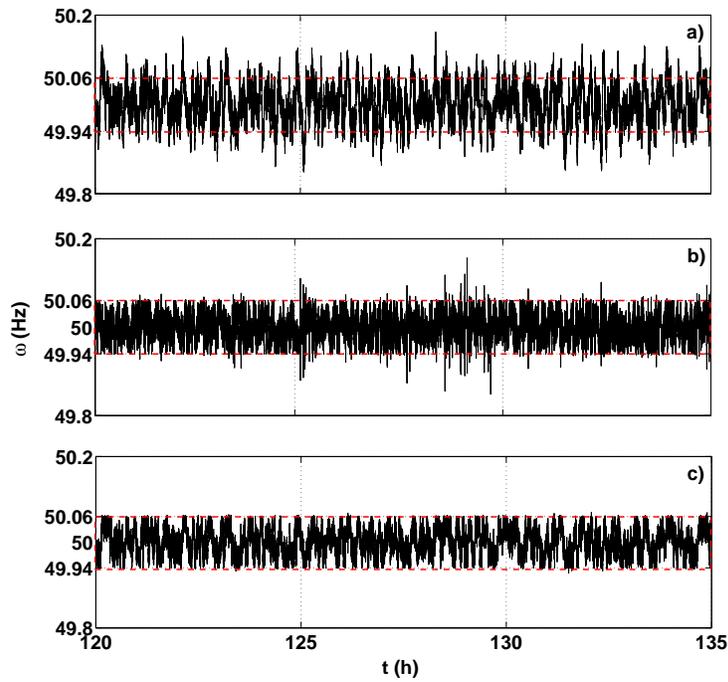}
 \caption{Time evolution of the electrical frequency: without DDC (a), with the basic DDC protocol (b) and with CeDDC when all devices communicate with each other (c). The red dashed line corresponds to 
 $\omega_{\rm R} \pm \epsilon_1$ which signals the boundaries in which pending tasks are recovered when DDC control is applied.}
\label{frequency_trajectory}
\end{figure}

\begin{figure}
\centering
   \includegraphics[width=0.6\linewidth] {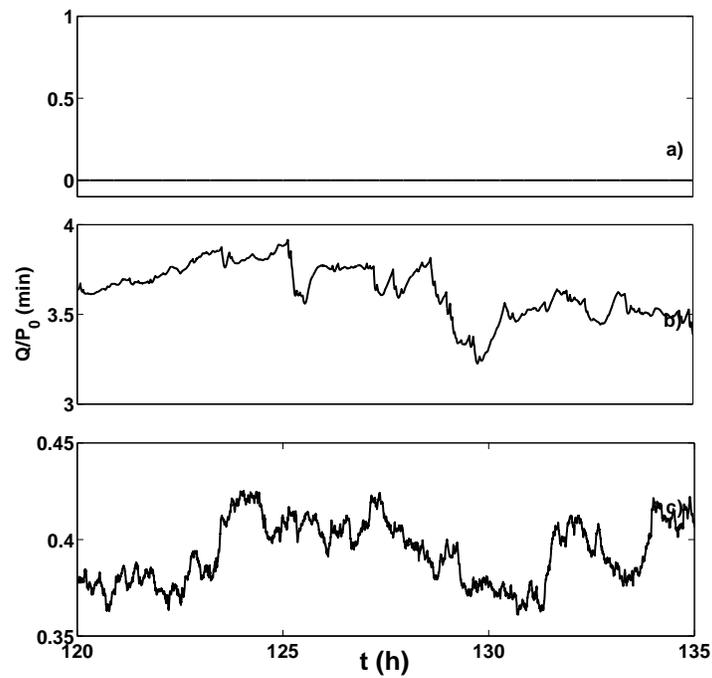}
 \caption{Time evolution of the average pending tasks per device scaled to $P_0$: without DDC (a), with the basic DDC protocol (b) and with CeDDC when all devices communicate with each other (c).}
\label{pending_tasks_trajectory}
\end{figure}

The dynamical behavior is illustrated in Figures \ref{load_trajectory} for the total electric load, \ref{frequency_trajectory} for the frequency and \ref{pending_tasks_trajectory} for the pending tasks.


Fig.~\ref{load_trajectory}a) shows a time trace for the total load obtained from a numerical simulation of the power plant equations (\ref{eq1})-(\ref{eq3}) with a fluctuating demand as given by equations (\ref{Pe})-(\ref{Ptotal}) without DDC, namely devices switch on and off when randomly schedule independently of the frequency conditions. This can be considered as the reference case. The corresponding time trace for the frequency is shown in Fig.~\ref{frequency_trajectory}a) and the time trace for the accumulated pending tasks in Fig.~\ref{pending_tasks_trajectory} a) (since DDC is not applied there are no pending tasks are accumulated). 

Panel b) in Figs.~\ref{load_trajectory}-\ref{pending_tasks_trajectory} show the corresponding time traces when DDC is applied to the same stochastic realization, namely all appliances intend to follow the same random schedule of switchings, but DDC postpones some tasks if frequency conditions are not suitable. There is a reduction of the fluctuations in the total load which translates in reducing the fluctuations in the frequency. As a direct consequence of DDC, now there is relevant number of pending tasks. Despite the overall reduction in the frequency fluctuations, the presence of a few large peaks in the frequency difference in \ref{frequency_trajectory}b) is noticeable. These rare events are triggered by the recovery of pending tasks. The effective rate of recovering pending tasks provided the frequency is in the right range is given by probability rate $p$ (or $q$) times the amount of pending tasks. This rate can be large if there are many tasks accumulated, and, as a consequence, there is a non negligible probability that the recovering of many pending tasks takes place simultaneously despite randomization. The simultaneous switching of many devices induces a large variation of the load which leads to large frequency fluctuations.

Panel c) in Figs.~\ref{load_trajectory}-\ref{pending_tasks_trajectory} show the corresponding time traces when CeDDC is applied to the same stochastic realization as before. Here we consider the case in which all devices communicate with each other. In this situation the curve for the total load can be seen as an intermediate case between the total load without DDC and with DDC. Fluctuations are reduced with respect to the case without control, but the change in the global profile is smaller than with basic DDC. This reflects the fact that CeDDC is less intrusive than basic DDC. Since CeDDC generates less reschedulings than basic DDC the amount of pending tasks (Fig.~\ref{pending_tasks_trajectory}) is almost $10$ times smaller than with DDC. As a consequence the large peaks in the frequency difference are no longer present and fluctuations are now kept within the range $[\omega_{\rm R}-\epsilon_1, \omega_{\rm R}+\epsilon_1]$. Curiously enough the variance  of the frequency fluctuations, it turns out to be $\sigma^2_{\omega} \approx 6.7 \times 10^{-4}$ s$^{-2}$ in the case of DDC and $\sigma^2_{\omega} \approx 9.2 \times 10^{-4}$ s$^{-2}$ in the case of CeDDC. The increase of the variance is a consequence of the changes in the frequency fluctuation pattern. For CeDDC there are periods of time in which the frequency fluctuates only in upper half of the range $[\omega_{\rm R},  \omega_{\rm R}+\epsilon_1]$ and periods of time in which the frequency fluctuates in the lower half range $[\omega_{\rm R}-\epsilon_1, \omega_{\rm R}]$. This sort of bimodal fluctuation pattern leads to a larger value for the variance than in the case of basic DDC. However since the fluctuations take place within the proper range the larger variance does not signal a deterioration of the performance in real situations. 

The change in the pattern of fluctuations is also reflected in the shape of the complementary cumulative distribution $R(\Delta\omega)$ shown in Fig.~\ref{cumulative_distribution}. For $\Delta \omega < \epsilon_1$, the green and blue lines corresponding to CeDDC are located above the red line corresponding to basic DDC without interaction and below the black line corresponding to the case without DDC (see inset). Thus CeDDC leads to a larger amount of frequency fluctuations at small scale than basic DDC, although compared to no DDC the amount of fluctuations is always greatly reduced. The relevance of the communication comes into play for fluctuations of size larger than $\epsilon_1$, for which the green and blue lines (CeDDC) are clearly below the red line (basic DDC).

\begin{figure}[ht]
\centering
   \includegraphics[width=0.6\textwidth] {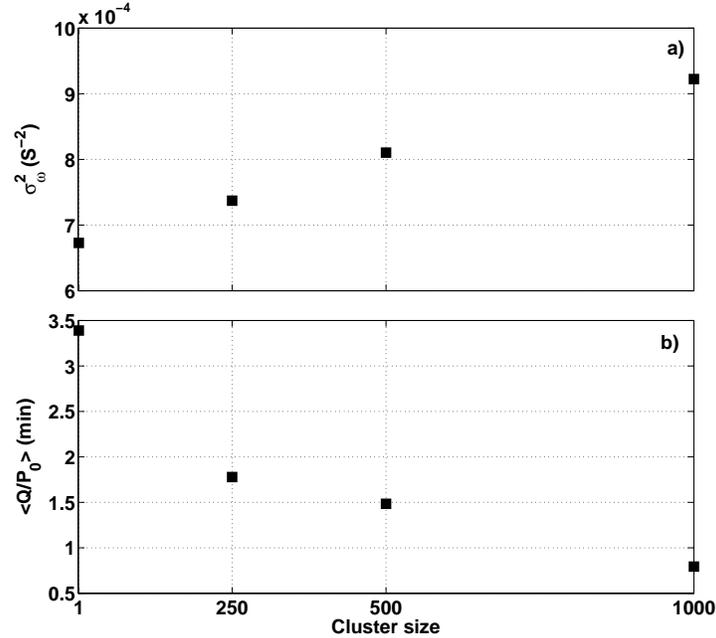}
 \caption{Dependence on the communication cluster size. (a) Variance of the frequency fluctuations and (b) average pending tasks per device scaled to $P_0$.}
\label{several_groups}
\end{figure}

We now address in more detail the effect of the size of the clusters in CeDDC. In particular we consider that interaction takes place within $4$ clusters of size $250$ or $2$ clusters of size $500$. Fig.~\ref{several_groups} compares the results for the dependence of the variance of the frequency fluctuations [panel a)] and for the pending tasks [panel b)] with the case of no basic DDC without interaction (cluster size $1$) and with the case of all to all interaction (cluster size $1000$). The variance of the frequency fluctuations increases with the group size due to the change in the dynamics discussed above. On the contrary, the average number of pending tasks decreases monotonously with the size of the group, as more opportunities for interchanging actions between devices are encountered. Very significant reductions of the pending tasks are already achieved for clusters of size $250$. As shown in Fig.~\ref{cumulative_distribution} for this cluster size, large frequency fluctuations are fully avoided in practise.

\begin{figure}[ht]
\centering
   \includegraphics[width=0.6\columnwidth] {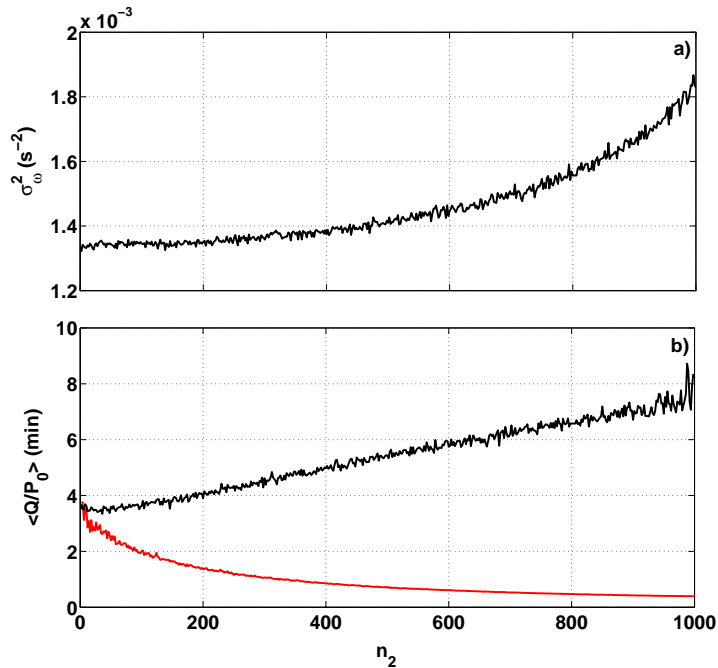} 
 \caption{Coexistence of DDC and CeDDC devices. (a) Variance of the frequency fluctuations $\sigma_{\omega}^2$ and (b) average pending tasks scaled to $P_0$ per basic DDC device (black line) and per CeDDC device (red line) as a function of the size of the CeDDC cluster $n_2$}
\label{DDC+CeDDC}
\end{figure}

Finally, we consider the coexistence of DDC and CeDDC, namely that out of the $N$ devices, $n_1$ devices follow the basic DDC protocol while $n_2=N-n-1$ interact among them. This can be viewed as having a interaction cluster of size $n_2$ and $n_1$ clusters integrated by a single device. In the following we identify each configuration by the pair of values $(n_1,n_2)$. Two particular cases have already been studied, namely $(N,0)$, which corresponds to basic DDC without interaction, and $(0,N)$ which corresponds to CeDDC with all-to-all interaction. For $n_1\ne0$ and $n_2\ne0$, interacting (CeDDC) and non-interacting (DDC) devices coexist in the system. We have performed several simulations with the same stochastic realization for the intended schedule of switchings, varying $n_1$. Fig.~\ref{DDC+CeDDC} a) shows the results for the variance of the frequency $\sigma_{\omega}^2$. As expected from the previous results the variance increases slightly with the number of interacting devices. Fig.~\ref{DDC+CeDDC} b) shows the results for the average pending tasks per device for non-interacting devices (black line) and for interacting devices (red line). For the interacting devices the number of pending tasks decreases monotonously with the size of the cluster $n_2$ (red line). A side effect of the communication in the smart devices is the increase of pending tasks in the non communicating DDC devices. This comes from the fact that CeDDC reduces the excursions of the frequency beyond the range $[\omega_{\rm R}-\epsilon_1, \omega_R+\epsilon_1]$ and, thus, the opportunities for basic DDC devices to recover pending tasks. This side effect could be avoided by choosing a value for the recovery threshold $\epsilon_1$ larger for communicating devices than for basic DDC without communications.

\begin{figure}
\centering
\includegraphics[width=0.6\linewidth]{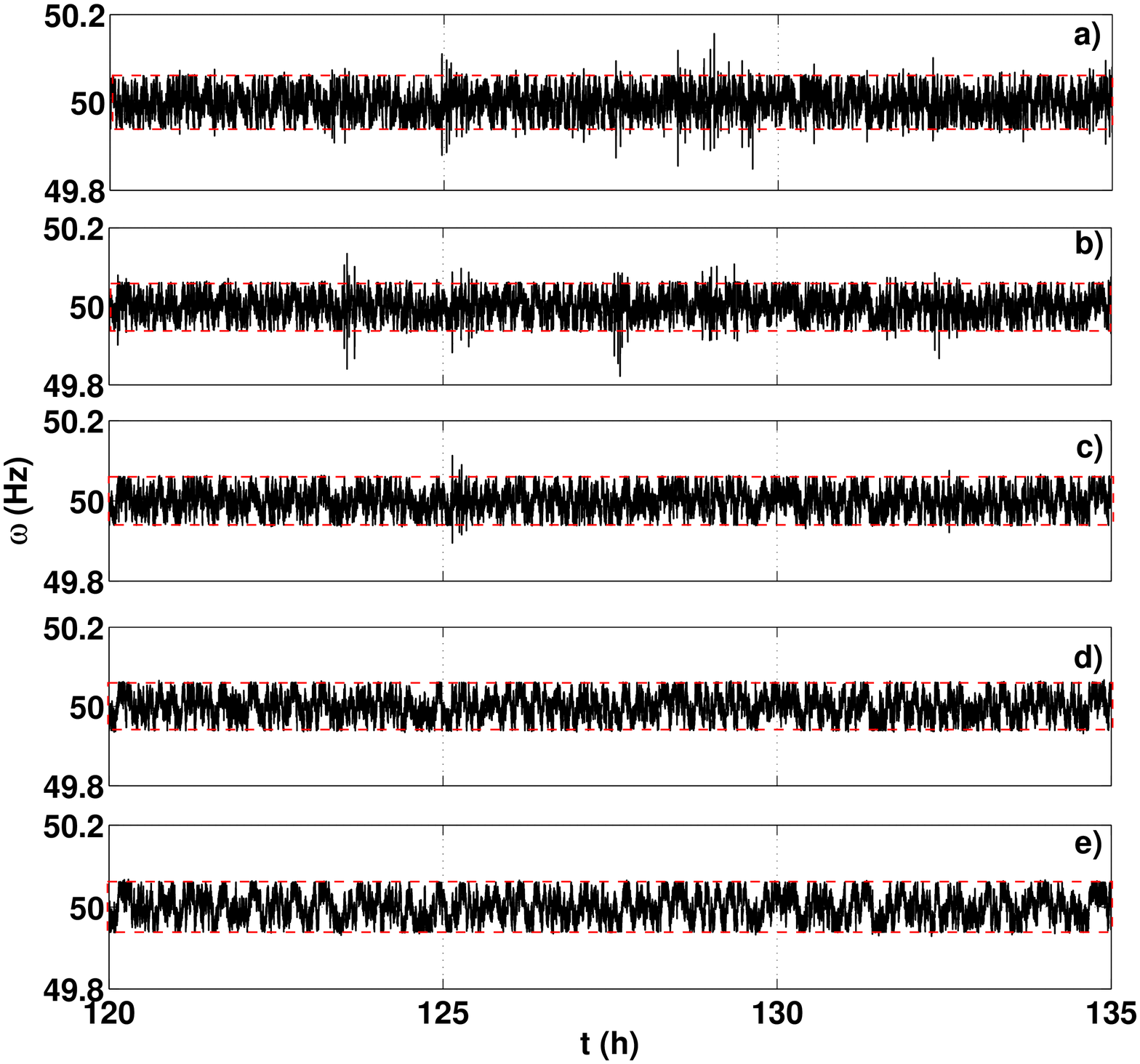}
\caption{Time evolution of the electrical frequency for coexisting DDC and CeDDC. We consider different amounts of non-connected and connected DDC devices $(n_1,n_2)$:  $(1000,0)$ (a); $(800,200)$ (b); $(500,500)$ (c); $(200,800)$ (d) and $(0,1000)$ (e). Red dashed lines correspond to  $\omega_{\rm R} \pm \epsilon_1$ which signals the boundaries in which pending tasks are recovered when DDC control is applied.}
\label{DDC+CeDDC_trajectory}
\end{figure}

\begin{figure}
\centering
\includegraphics[width=0.6\linewidth] {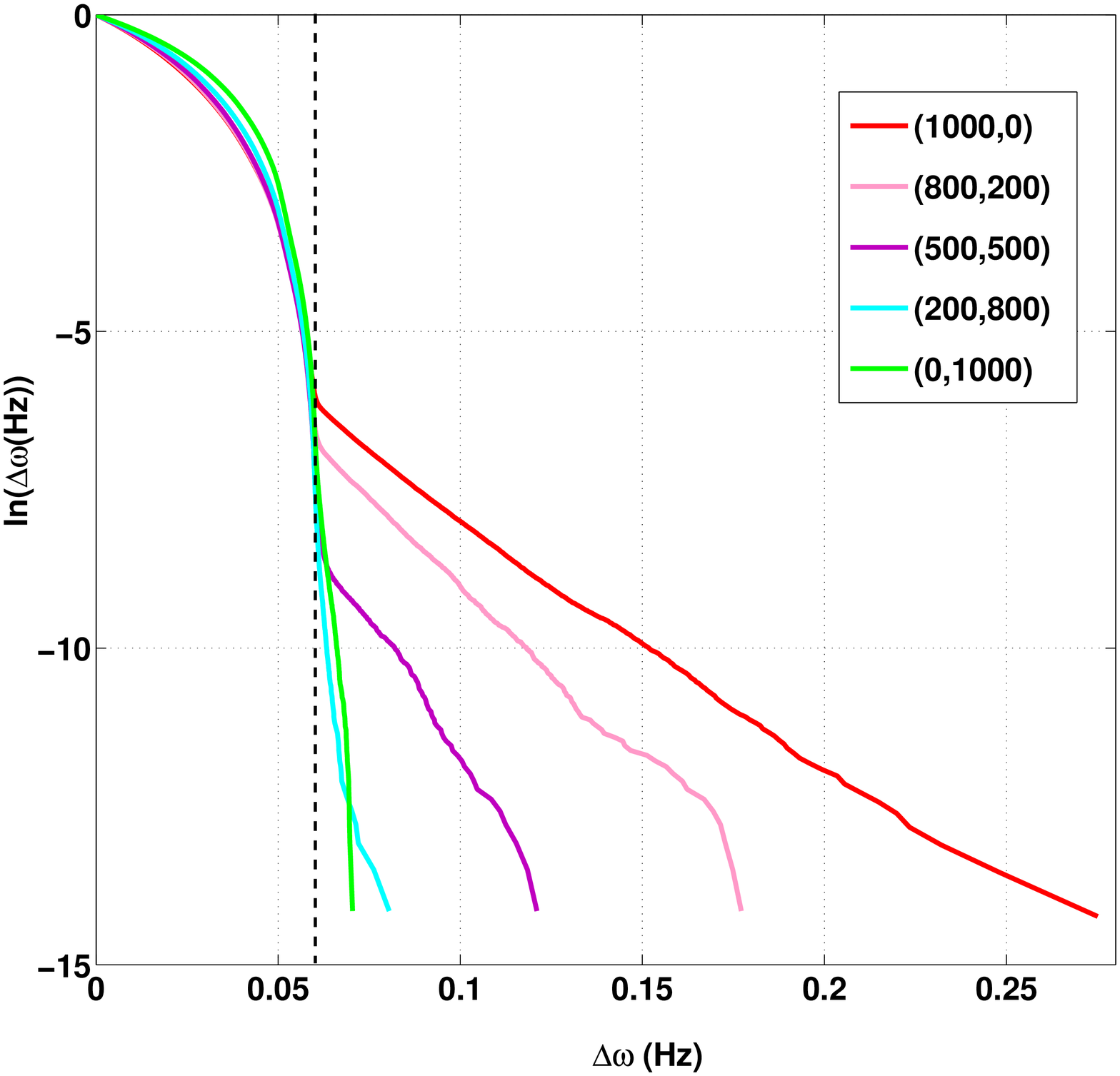} 
\caption{Estimated cumulative probability of having a frequency fluctuation of size larger than $\Delta \omega$, $R(\Delta \omega)$. Colors correspond to different amounts of non-connected and connected devices $(n_1,n_2)$: $(1000,0)$ red, $(800,200)$ brown, $(500,500)$ purple, $(200,800)$ blue and $(0,1000)$ green.}
\label{cumulative_distribution_n2}
\end{figure}

Figure~\ref{DDC+CeDDC_trajectory} shows the time trace for the frequency for several values of $(n_1,n-2)$. As it can be seen already with a cluster of $n_2=200$ communicating devices the probability to observe large frequency fluctuations is greatly reduced. Increasing $n_2$ leads to further reduction of this probability. Thus, including a fraction of devices which communicate each other can improve the overall stability of the power grid with a significant reduction of the probability to have large frequency fluctuations, despite the fact that non-communicating devices are now accumulating more pending tasks than before. This illustrative results of the dynamics can be quantified by looking at the the probability to have a frequency fluctuation of size larger than $\Delta \omega$, $R(\Delta \omega$), which is plotted in Fig.~\ref{cumulative_distribution_n2}. As in Fig.~\ref{cumulative_distribution} the red line corresponds to the case of basic DDC without control while the green line corresponds to the case of CeDDC with all-to-all interaction. Other lines correspond to the coexistence of non-communicating and communicating devices in the same ratios as in Fig.~\ref{DDC+CeDDC_trajectory}. As the number of devices with communication, $n_2$, increases the tails of the distribution are shortened. In fact even with only $20\%$ of devices communicating among them there is a significant impact in the probability to have large size frequency fluctuations. As an example the probability of fluctuations larger than $0.1$ is reduced by a factor $10$.  ({\bf Verificar si la escala logartimica es con el log en base 10 o natural}). 

\section*{Discussion}

We have introduced a DDC algorithm which allows for communication between devices which are part of a given cluster, in such a way that devices can switch on (off) even if the frequency is outside the suitable range provided another device within the cluster has switched off (on). This communication-enhanced DDC protocol is less intrusive than basic DDC, allowing more switches and reducing in a very significant way (up to 90\%) the accumulated pending tasks. One of the major advantages of the protocol introduced here is that the probability of large scale fluctuations in the electrical frequency is strongly reduced and can even be practically suppressed. Having a non-negligible probability for large size fluctuations is one of the drawbacks of simple DDC algorithm that can no be avoided by tweaking parameters and which is relevant for practical implementations since large frequency fluctuations can trigger a failure of the power grid at large scale.  

Large scale frequency fluctuations are fully avoided for all-to-all interaction. However an ideal all-to-all communication would require nation-wide internet communication and probably would not be practical. Nevertheless technologies as power line communication (PLC) could allow devices to interchange information over a certain range of distances. Thus an scenario of clusters of devices seems more plausible while still providing a very significant reduction of the probability of large size frequency fluctuations. Communication can be implemented through a hub such as a household energy management or by direct communication among the devices of a cluster.



\bibliography{interaction}

\section*{Acknowledgements}

We acknowledge financial support from Ministerio de Econom\'{\i}a, Industria y Competividad (MINEICO, Spain), Agencia Estatal de Investigaci\'on (AEI, Spain) and Fondo Europeo de Desarrollo Regional under Project ESoTECoS, grant numbers: FIS2015-63628-C2-1-R (MINEICO/AEI/FEDER,UE) and FIS2015-63628-C2-2-R (MINEICO/AEI/FEDER,UE). E.B.T.T. also acknowledges the fellowship FIS2015-63628-CZ-Z-R under the FPI program of MINEICO, Spain.

\section*{Author contributions statement}

E.B.T.T. has performed the numerical calculations and prepared the figures. All authors have discussed the model, analyzed the results and prepared the manuscript. 

\section*{Additional information}

The authors declare no competing interests.


\end{document}